\documentclass{osa-article}

\journal{osajournal}

\usepackage{tabularx}
\usepackage{multirow}
\begin{document}

\title{Two-dimensional extreme skin depth engineering for CMOS photonics}

\author{Matthew van Niekerk\authormark{1,*}, 
Saman Jahani\authormark{2,3},
Justin Bickford\authormark{4},
Pak Cho\authormark{4},
Stephen Anderson\authormark{4,5},
Gerald Leake\authormark{6},
Daniel Coleman\authormark{6},
Michael L. Fanto\authormark{1,7},
Christopher C. Tison\authormark{7},
Gregory A. Howland\authormark{1,8},
Zubin Jacob\authormark{2},
and Stefan F. Preble\authormark{1}}

\address{\authormark{1} Microsystems Engineering, Rochester Institute of Technology, Rochester, NY 14623, USA \\
\authormark{2} School of Electrical and Computer Engineering and Birck Nanotechnology Center, Purdue University, West Lafayette, IN 47907, USA \\
\authormark{3} Department of Electrical Engineering, California Institute of Technology, Pasadena, CA, 91125, USA \\
\authormark{4} U.S Army Research Laboratory , Adelphi, MD 20783 USA \\
\authormark{5} Electrical, Computer and Systems Engineering Department, Rensselaer Polytechnic Institute, Troy, NY 12180, USA \\
\authormark{6} State University of New York Polytechnic Institute, Albany, NY 12203, USA \\
\authormark{7} Air Force Research Laboratory, Rome, NY 13441, USA \\
\authormark{8} School of Physics and Astronomy, Rochester Institute of Technology, Rochester, NY 14623, USA
}
\email{\authormark{*}mv7146@rit.edu} 



\begin{abstract}
Extreme skin depth engineering (\emph{e-skid}) can be applied to integrated photonics to manipulate the evanescent field of a waveguide.
Here we demonstrate that \emph{e-skid} can be implemented in two directions in order to deterministically engineer the evanescent wave allowing for dense integration with enhanced functionalities. 
In particular, by increasing the skin depth, we enable the creation of large gap, \emph{bendless} directional couplers  with large operational bandwidth.  Here we experimentally validate two-dimensional \emph{e-skid} for integrated photonics in a CMOS photonics foundry and demonstrate strong coupling with a gap of $1.44$ $\mu$m.
\end{abstract}

\section{Evanescent Waves in Silicon Photonics}
Evanescent waves in silicon photonic waveguides have the propensity to cause parasitic optical crosstalk.
In traditional photonic circuits, design strategies must consider minimum separation distances between any closely spaced waveguides to prevent unwanted coupling \cite{chrostowski2015silicon}. 
This problem is inhibiting for many photonic circuits due to cost and size constraints.
Many efforts have been made to overcome these issues, battling size constraints by employing inverse design \cite{piggott2015inverse,shen2015integrated} or implementing metamaterials to increase performance \cite{jahani2016all,luque2018tilted, staude2017metamaterial}.
\par 
Recent work introduced a new, metamaterial paradigm for waveguiding that fundamentally suppresses coupling between waveguides \cite{jahani2018controlling}. 
In this approach a subwavelength, multi-layer cladding is placed in plane and in parallel with the waveguide, decreasing the skin depth of the fundamental transverse-electric (TE) mode's evanescent field. 
The concept is called \emph{extreme skin depth engineering}, or \emph{e-skid}.
\emph{E-skid} has been employed as cross-talk suppresion \cite{jahani2018controlling,mia2020exceptional},
and for high performance polarization splitting \cite{xu2019anisotropic,chen2019high}.
The \emph{e-skid} features are created in the same processing step as the waveguide itself, allowing this to be an innate no-cost addition to any design. 
The addition of these features can reduce the crosstalk between waveguides by more than three orders of magnitude, which will dramatically reduce the photonic design footprint. \cite{jahani2018controlling}. 
\par 
Here we expand on this work by using \emph{e-skid} in \emph{two directions}. 
Using both a parallel \cite{jahani2018controlling} and perpendicular cladding we can engineer the coupling between waveguides throughout a photonic circuit. 
Specifically, a perpendicular eskid cladding can increase coupling by up to four orders of magnitude.
Using this, we design a large gap, bendless directional coupler that operates over a large bandwidth ($\geq 40$ nm).
Employing \emph{e-skid} techniques to traditional photonic components allows an immediate decrease in overall system footprint, not just limited to waveguide routing.
Finally, we demonstrate these \emph{two direction} \emph{e-skid} large gap directional couplers in a complementary metal-oxide semiconductor (CMOS) photonic platform, thereby affirming the manufacturability of \emph{e-skid} components and integration with foundry offerings.
\begin{figure}
    \centering
    \includegraphics[width = \textwidth]{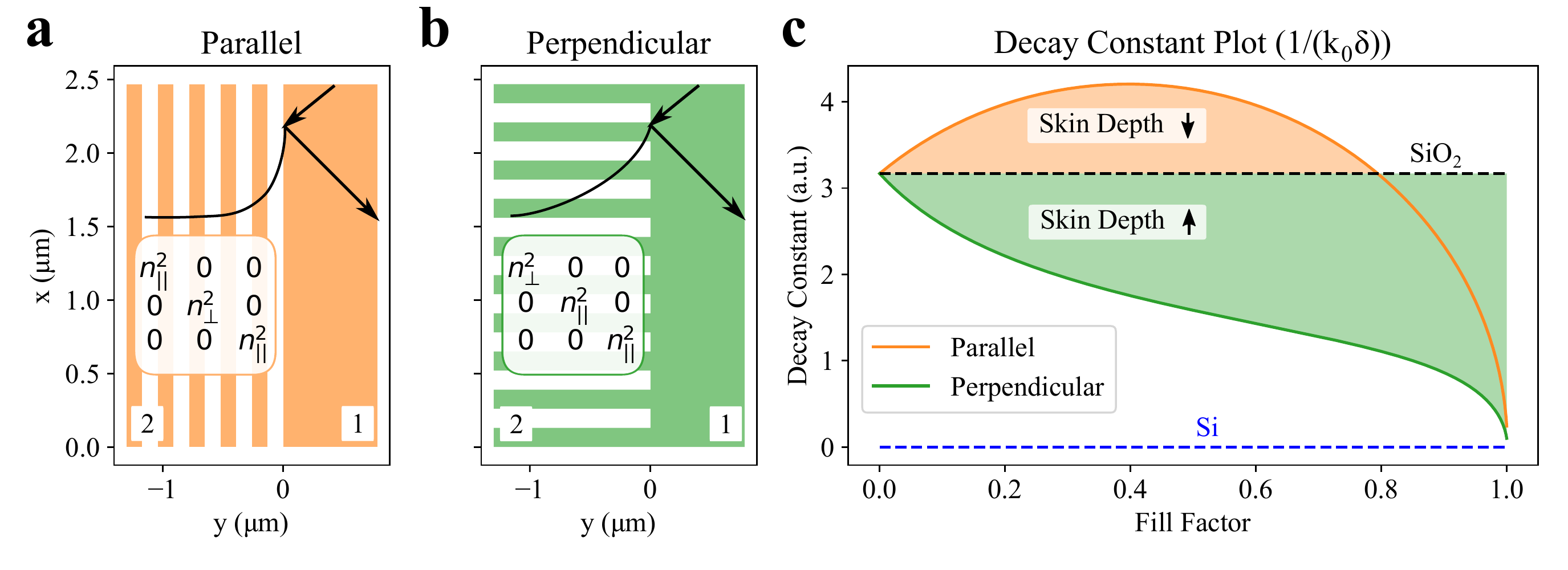}
    \caption{(a) Here we demonstrate the parallel oriented subwavelength, multi-layer cladding. The anisotropic permittivity tensor is displayed over the cladding, which follows the Rytov relations for each direction.  (b) The perpendicular cladding essentially swaps the $xx$ and $yy$ components of the permittivity tensor in (a). The incident wave is reflected (shown by the two arrows in medium one) and the evanescent wave is strongly decaying in medium two in (a) and weakly decaying in (b). (c) A plot of Eq. \ref{eq: reduced_sd} for the two different cladding strategies. We see that decay is increased over SiO$_2$ for most of the fill factors of the parallel case, whereas we can see a variable decrease in decay by almost the whole scale between the two materials in the perpendicular cladding.}
    \label{fig: cladding}
\end{figure}
\section{Theory}
\subsection{\emph{E-skid} in Two Directions}
Consider two media with index of refraction $n_1,n_2$.
When an incoming wave from $n_1$ meets the boundary at $n_2$ and the angle is greater than the critical angle $\theta_i < \theta_c = \operatorname{sin}^{-1}(n_2/n_1)$ an evanescent wave is formed in the second medium. 
This wave does not carry power across the boundary; it exponentially decays into the second medium \cite{hecht2002optics}.  
\emph{E-skid} allows us to tune the decay constant of this evanescent wave by introducing subwavelength, periodic structures that transform wave's (specifically, a polarized wave's) momentum \cite{Jahani:14,jahani2015photonic}. 
These features change the second medium from an isotropic material to an anisotropic metamaterial.
The anisotropy here refers to the permittivity values of the dielectric tensor of the material (where we are assuming that the permittivity can be defined $\epsilon_r = n^2$). 
For deep subwavelength features, these component values are defined by the Rytov relations \cite{rytov1956electromagnetic,cheben2018subwavelength}:
\begin{equation}\label{eq: Rytov 1}
n_{\parallel}^2 = n_{1}^2 \rho + n_{2}^2 (1-\rho),
\end{equation} 
\begin{equation}\label{eq: Rytov 2}
n_{\bot}^{-2} =  n_{1}^{-2} \rho + n_{2}^{-2} (1-\rho),
\end{equation} 
where $n_{1},n_{2}$ are the indices of the first and second medium, respectively, and $\rho$ is the fill factor. 
The parallel component, $n_{\parallel}$, is defined in the direction parallel to the periodic structure's orientation, and the perpendicular component, $n_{\bot}$, is oriented perpendicular to the periodic structure. 
For features that are deep-subwavelength, these relations demonstrate how the second material transforms from isotropic to an anisotropic metamaterial, however any subwavelength structures will exhibit anisotropy albeit without these neat relations.
\begin{figure}
    \centering
    \includegraphics[width=\textwidth]{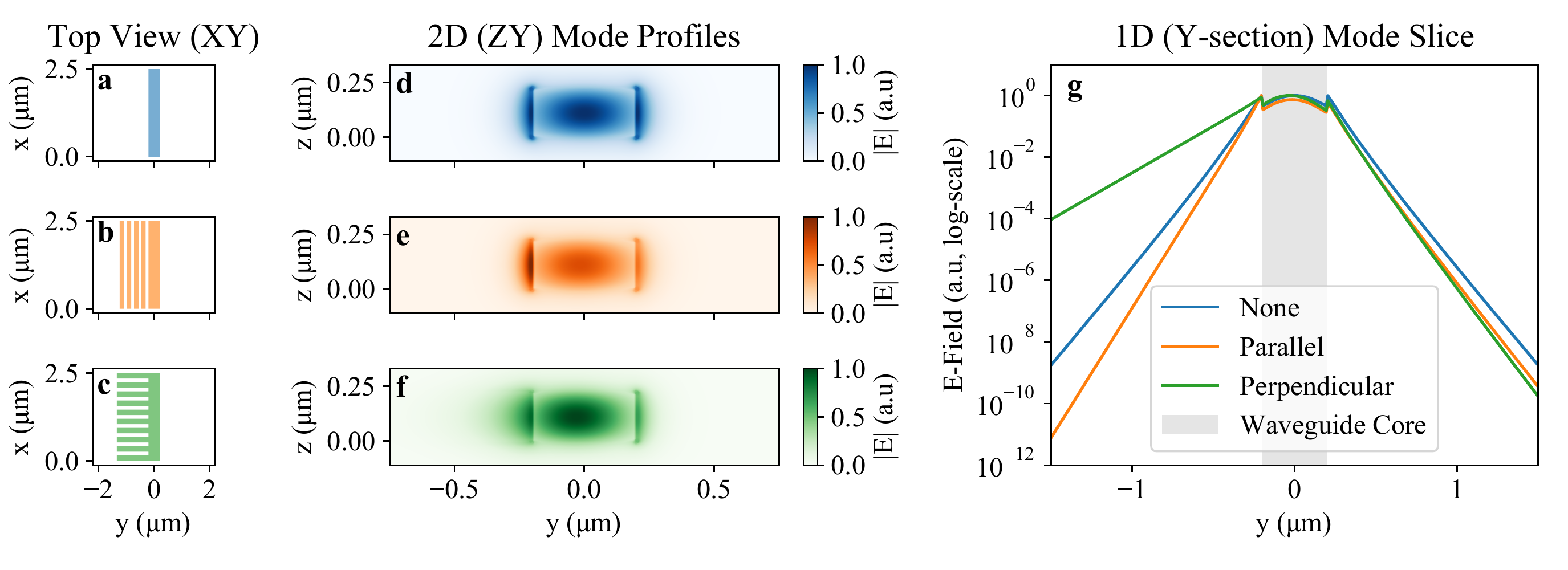}
    \caption{A comparison of the different cladding strategies discussed, with the note that they have a common cladding on the right hand side (normal isotropic SiO$_2$). (a,b,c) Here we have the top-down perspective of the waveguides, which shows the cladding for the none, parallel and perpendicular structures on the left hand side, respectively. (d,e,f) The ZY plane cross section mode profiles corresponding to the cladding diagrams from (a,b,c), where the width of the waveguides is $400$ nm and the height is $220$ nm. (g) A center slice through each of the mode profiles, which demonstrate, on a log-linear scale, the amount of control we can impose on the evanescent wave with these structures. Simulations were done with an anisotropic material following Rytov relations (Eqs. \ref{eq: Rytov 1},\ref{eq: Rytov 2}), where the core waveguide width was $400$ nm, and fill factor was 0.6 for both orientations.}
    \label{fig: modes}
\end{figure}
The key result of the \emph{e-skid} derivation leverages this anisotropy for the evanescent wave, which is characterized by the decay constant, $\beta$:
\begin{equation}\label{eq: reduced_sd}
\beta(\theta_i;\rho) = \frac{1}{\delta(\theta_i)} = k_0\frac{n_{2x}(\rho)}{n_{2y}(\rho)}\sqrt{n_1^2 sin^2(\theta_i) - n_{2y}(\rho)^2}.
\end{equation}
where $k_0$ is the wavevector and $\theta_i$ is the angle of the incident wave to the boundary (we assume paraxial $\theta_i \approx \pi/2$) \cite{Jahani:14}.
The decay constant is now subject to a degree of variable tunability ($\rho$), allowing for control of the evanescent wave \cite{jahani2018controlling,Jahani:14}.
\par
In Fig. \ref{fig: cladding} (a), we show the dielectric tensor for the \emph{e-skid} structure, where the periodicity of the subwavelength features is parallel to the boundary ($y = 0$). 
In this orientation, the diagonal components of the second material become $[n_{2x}^2,n_{2y}^2,n_{2z}^2] = [n_{\parallel}^2,n_{\bot}^2,n_{\parallel}^2]$ in accordance with the Rytov relations (Eqs. \ref{eq: Rytov 1}, \ref{eq: Rytov 2}). 
This structure will increase the decay constant of the evanescent wave, thereby decreasing the skin depth \cite{jahani2018controlling}.
Without loss of generality, we recognize that we can rotate the optical axis by rotating the subwavelength features and realize \emph{e-skid} in a \emph{second} direction. 
Due to the direction dependency outlined by the Rytov relations, when we rotate the periodicity of the features, we effectively swap the $xx$ and $yy$ components of the dielectric tensor of the parallel cladding such that we now see $[n_{2x}^2,n_{2y}^2,n_{2z}^2] = [n_{\bot}^2,n_{\parallel}^2,n_{\parallel}^2]$ (Fig. \ref{fig: cladding} (b)). 
The values of $n_{2x},n_{2y}$ in Eq. \ref{eq: reduced_sd} control the decay constant, and by rotating the periodic structure we are able to dictate an increase or decrease.
\par 
We populated Eq. \ref{eq: reduced_sd} with the new dielectric tensor values outlined in Fig. \ref{fig: cladding} (a,b) such that we show in Fig. \ref{fig: cladding} (c) the full range of decay constant tunability of \emph{e-skid} in two directions. 
Fig. \ref{fig: cladding} (c) shows clearly that both decreasing and increasing skin depth can be achieved by the parallel features, however applying this to CMOS photonics manufacturing, we generally omit the higher and lower fill factors due to resolution constraints. \cite{smith1998microlithography}.
We used a material platform consistent with CMOS photonics in (c), such that material one is silicon (Si) and material two is silicon dioxide (SiO$_2$), however, this is true for any optical material combination as long as $n_1>n_2$.
\subsection{\emph{E-skid} in Two Directions in Waveguides}
Optical waveguiding is not fully described by the simple electromagnetic wave-at-a-boundary example above. 
While it lends intuition, we must find the electromagnetic mode of the entire structure to get a clear picture of this effect.
We used a commercial finite difference eigenmode (FDE) solver to simulate three specific types of waveguides to demonstrate \emph{e-skid} in two directions \cite{lumerical}. 
Fig. \ref{fig: modes} (a) shows a top view of a single-mode strip waveguide, where the propagation is in the $\hat{x}$ direction. 
Next to the strip waveguide, Fig. \ref{fig: modes} (d) shows a 2D profile of the fundamental TE propagating mode. 
We introduce the wave supressing \emph{e-skid} features on one side of the waveguide in Fig. \ref{fig: modes} (b) and show the corresponding 2D mode in (e).
Finally, we introduce the wave enhancing \emph{e-skid} features in Fig. \ref{fig: modes} (c) and the corresponding 2D mode in (f). 
We compiled the cross sections of all three modes in Fig. \ref{fig: modes} (g) to demonstrate the effect of the features on the evanescent wave of the mode. 
Fig. \ref{fig: modes} (g) clearly demonstrates, with a log-scale in $y$, that the decaying wave outside of the center of the waveguide is suppressed by the parallel features, and greatly enhanced by the perpendicular features.
\begin{figure}[ht!]
    \centering
    \includegraphics[width = \textwidth]{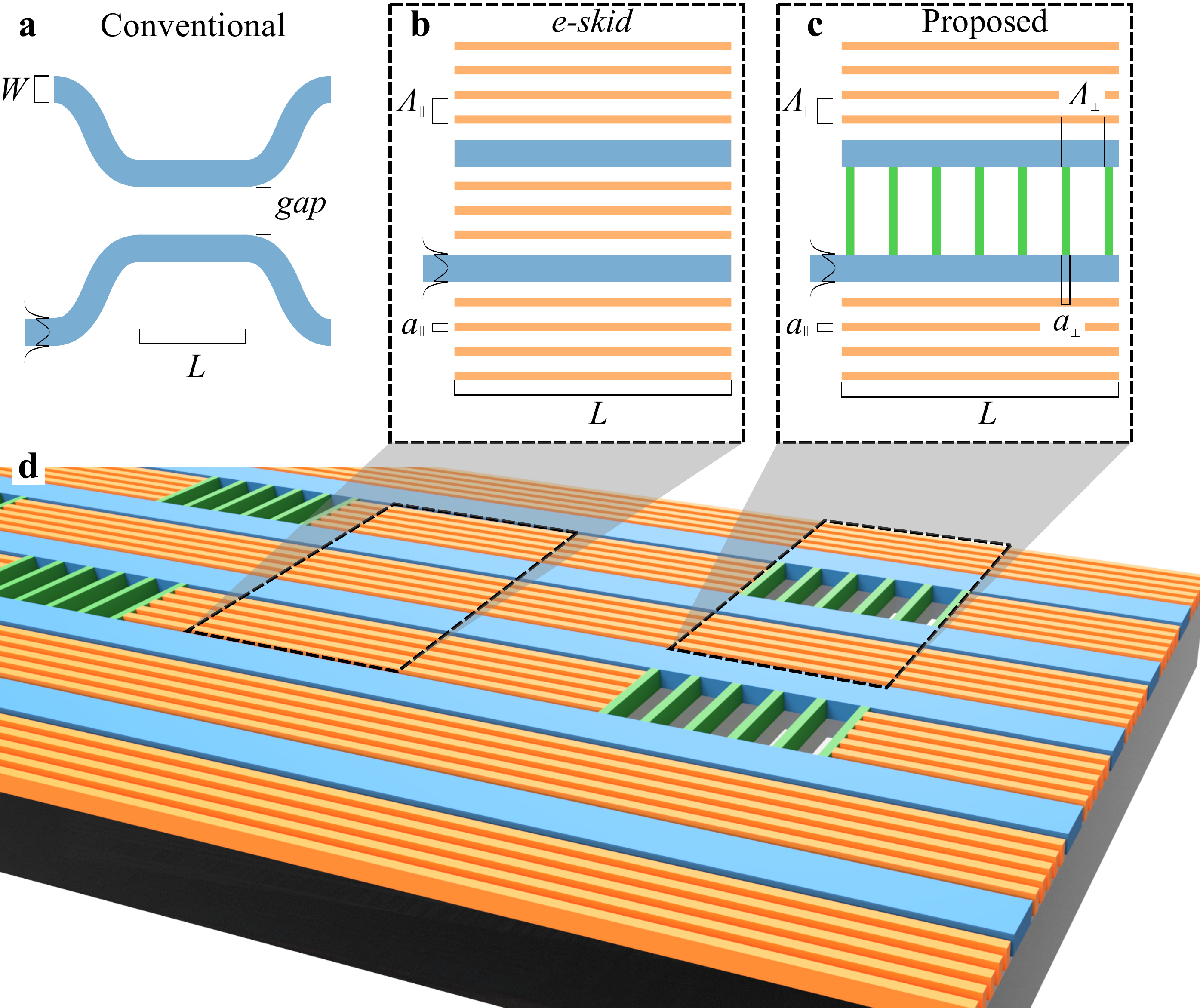}
    \caption{(a) A conventional directional coupler with a coupling region characterized by the gap between waveguides and the length of the parallel section. (b) The \emph{e-skid} platform discussed in \cite{jahani2018controlling}, where the period ($\Lambda_{\parallel}$) and silicon fill ($a_{\parallel}$) characterize the subwavelength features. (c) Our directional coupler which leverages the enhancing \emph{e-skid} features in the coupling region, where the features outside the coupling region are the same as (b) and where the period ($\Lambda_{\bot}$) and silicon fill ($a_{\bot}$) characterize the subwavelength features in the coupling region.
    (d) An example of an integrated photonic circuit implementing two-dimensional \emph{e-skid}. Note the circuit maintains the size reduction of \emph{e-skid} coupled with large gap, bendless couplers. Colors throughout indicate the photonic waveguides (blue), parallel (orange) and perpendicular (green) \emph{e-skid} features (where all are the same material - namely silicon), and the base is the buried oxide (black). }
    \label{fig: design}
\end{figure}
\section{Large Gap, Bendless Directional Coupler Design}
We propose and demonstrate a new coupler that leverages \emph{e-skid} in two directions to create coupling in desired regions.
In a traditional integrated photonic platform, a directional coupler is created by bending two waveguides close to each other. (Fig. \ref{fig: design}(a)).
The waveguides must otherwise be kept far apart in other parts of a circuit in order to avoid unwanted coupling, limiting the circuit density.
By using \emph{e-skid} with parallel subwavelength features (Fig. \ref{fig: design}(b)), that suppress coupling, we overcome this limitation and keep two waveguides within close proximity, with negligible coupling.  
Furthermore, when the design with \emph{e-skid} needs coupling, we show that the introduction of perpendicular subwavelength features in the coupling region, as are seen in Fig. \ref{fig: design}(c), will significantly enhance coupling. 
These features have tunable variables (i) period ($\Lambda_{\bot}$) and (ii) fill factor ($\rho_{\bot}$) which directly tune the amount of coupling experienced. 
We introduce two-dimensional \emph{e-skid} as a way to leverage the size reduction offered by the parallel features with the addition of the perpendicular features to create practical circuits as seen in Fig. \ref{fig: design} (d).
\subsection{Coupled Modes for \emph{E-skid}}
The evanescent wave of the mode, even though it carries no power across the boundary, causes coupling between parallel guides if the overlap between the evanescent waves of supported modes is large enough \cite{Huang:94}. 
From coupled mode theory \cite{Huang:94}, we define the power in the bar and cross ports as
\begin{equation}\label{eq: bar}
P_{bar}(L) = P_0 \operatorname{cos}^2\left(\kappa L\right) = P_0 \operatorname{cos}^2\left(\frac{\pi}{2}\frac{L}{L_x}\right),
\end{equation} 
\begin{equation}\label{eq: cross}
P_{cross}(L) =P_0 \operatorname{sin}^2\left(\kappa L\right) =  P_0 \operatorname{sin}^2\left(\frac{\pi}{2}\frac{L}{L_x}\right),
\end{equation} 
for $L$ as the coupling length, $P_0$ as the injected power, $\kappa$ as the coupling coefficient, $L_x$ as the crossover length such that $\kappa = \pi/(2 L_x)$, and bar and cross refer to the light remaining in the injected waveguide or transitioning to the other waveguide, respectively.
The crossover length is defined so that when $L = L_x$, there is complete power transfer from waveguide one to two. 
\par
This approach allows for an intuitive understanding of the device.
The crossover length is given as
\begin{equation}\label{eq: Lx}
L_x = \frac{\lambda}{2(n_{even,\lambda} - n_{odd,\lambda})},
\end{equation}
where $\lambda$ is the free space wavelength and $n_{even,\lambda}, n_{odd,\lambda}$ are the effective indices of the even and odd modes, respectively \cite{chrostowski2015silicon}.
The field of the odd mode is antisymmetric across the coupling region and it remains generally unaffected by symmetric features there \cite{halir2012colorless}.
However, by introducing features into the coupling region the even mode is affected, thereby enabling dispersion engineering of the directional coupler - specifically, controlling the directional couplers' optical bandwidth.  \cite{halir2012colorless}.
By crafting $L_x$, we can dictate how the device performs according to Eqs. \ref{eq: bar},\ref{eq: cross}. 
Essentially, if we make the slope of $L_x$ as flat as possible over a span of $\lambda$, we ensure a useful operating bandwidth (e.g. a 3 dB coupler) is preserved for that span.
Our design is fundamentally different than \cite{halir2012colorless} due to the structural asymmetry, which encourages coupling, and the higher fill factor.
These parameters allow us to create a directional coupler with more than an order of magnitude shorter crossover length in comparison, at the penalty of reduced operating bandwidth.
\par
\begin{figure}
    \centering
    \includegraphics[width = \textwidth]{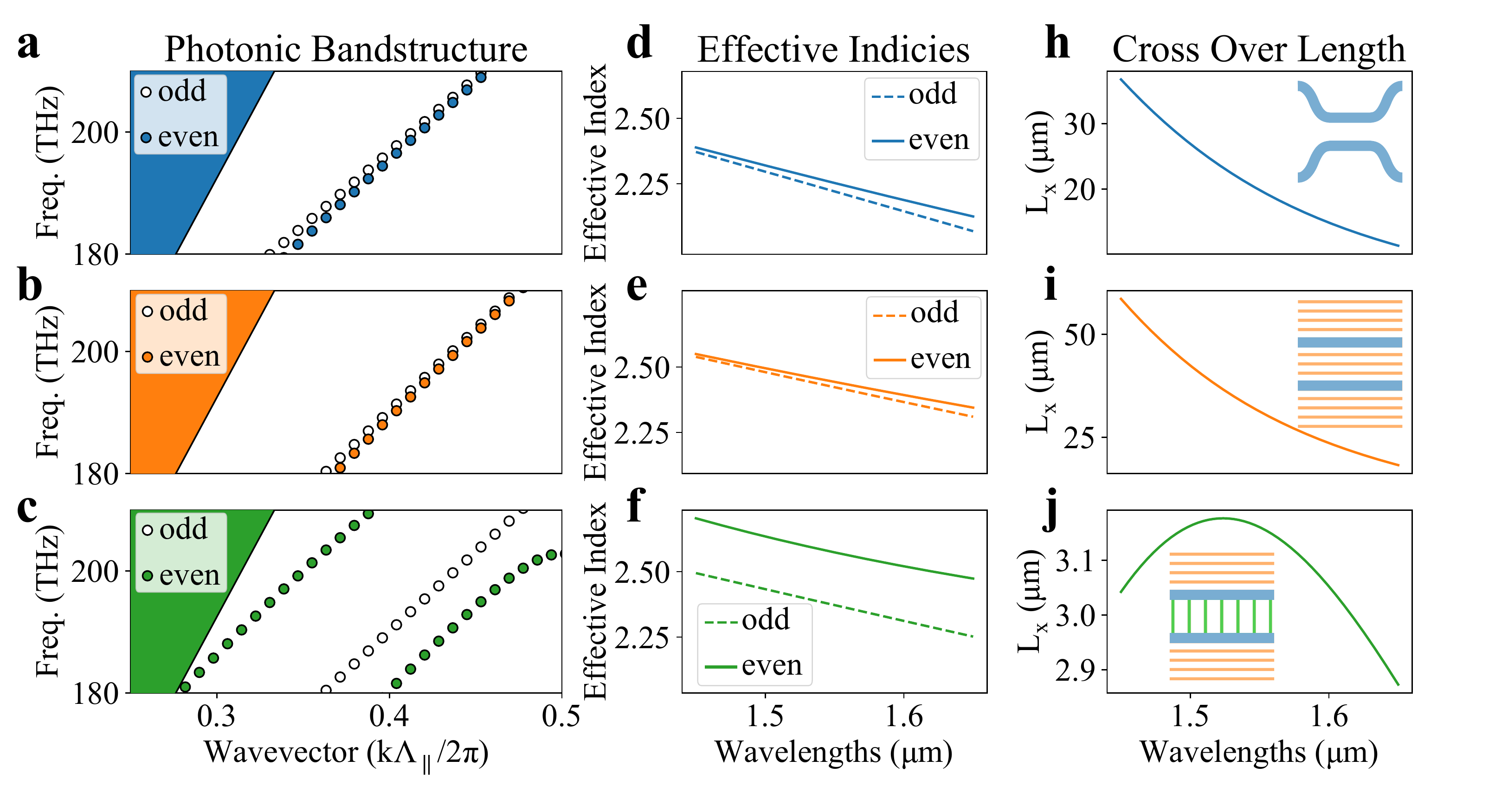}
    \caption{(a,b,c) The photonic bandstructures of the conventional, \emph{e-skid} and large gap directional couplers from Fig. \ref{fig: design} (a,b,c), respectively.
    (d,e,f) Extracted effective indices of the bandstructures from (a,b,c), respectively.
    (g,h,i) The crossover length, calculated from Eq. \ref{eq: Lx}. The insets of (g,h,i) show the device diagrams for each type of coupler.
    These devices were all simulated with the same gap to illustrate the effects on the same scale.
    In practice, the gap is limited by (i) the fabrication process, (ii) the circuit application, and (iii) the length of the waveguides. Therefore, the gap is often larger than shown here.
    The \emph{e-skid} design parameters were: gap $=270$ nm, $\Lambda_{\parallel} = 50$ nm, $\rho_{\parallel} = 50\%$, $\Lambda_{\bot} = 270$ nm, $\rho_{\bot} = 50\%$, and $W = 400$ nm. }
    \label{fig: bandstructure}
\end{figure}

In order to demonstrate the effect of dispersion engineering, we begin by simulating the photonic bandstructure of the directional coupler in a full wave 3D frequency time frequency domain (FDTD) solver with Bloch-periodic boundary conditions \cite{lumerical}.
Fig. \ref{fig: bandstructure} (a,b,c) shows the photonic bandstructure of a traditional, \emph{e-skid} and large gap directional coupler. 
These directional couplers are fundamentally different from photonic crystals as they are not designed to work in the band gap, instead these subwavelength features allow for low loss propagation through the periodic structures \cite{cheben2018subwavelength}.
The traditional and \emph{e-skid} couplers exhibit similar bandstructures, but the large gap coupler's even mode is approaching the band edge just above $200$ THz ($1500$ nm). 
Because the even mode is near the band edge, dispersion is increased, which allows for flexibility in tuning the behavior.
From the bandstructures, we extract the dispersive effective index of both the fundamental even and odd supermodes (Fig. \ref{fig: bandstructure} (d,e,f)).
The effective indices exhibit a similar characteristic shape to their corresponding bandstructures.
The \emph{e-skid} coupler brings the even mode effective index much closer to the odd mode in comparison with the traditional coupler, and the large gap coupler has increased the difference between the two effective indices. 
Fig. \ref{fig: bandstructure} (g,h,i) show the crossover length given the corresponding effective indices. 
The traditional coupler and \emph{e-skid} behave as expected, with an increase in crossover length for the latter.
The large gap coupler exhibits a dramatically reduced crossover length, and in relation to dispersion engineering, a completely different shape.
It is important to note that the large gap directional coupler does support a higher-order mode (Fig. \ref{fig: bandstructure} (c)), however the coupling efficiency extracted from a modal overlap integral between the fundamental and the first higher-order mode is $<8\%$ over the wavelength span for some designs, according to our 3D FDTD simulations \cite{lumerical}. 
While $8\%$ is not insignificant, tweaking our parameters (specifically $\rho_{\bot}$ and $\Lambda_{\bot}$) can reduce this coupling efficiency into higher-order modes, increasing device performance \cite{halir2012colorless}. 
For this experiment, the primary design choices were dictated from a perspective of manufacturability.
In the future, small decreases to $\rho_{\bot}$ and $\Lambda_{\bot}$ will result in higher performing couplers verified by our 3D FDTD simulations \cite{lumerical} and prior work \cite{halir2012colorless}. 
Additionally, careful consideration of tapering, which is not investigated in this work, can also mitigate the excitation of higher-order modes \cite{cheben2018subwavelength,halir2015waveguide}.
\subsection{Device Design}
\begin{figure}
    \centering
    \includegraphics[width = \textwidth]{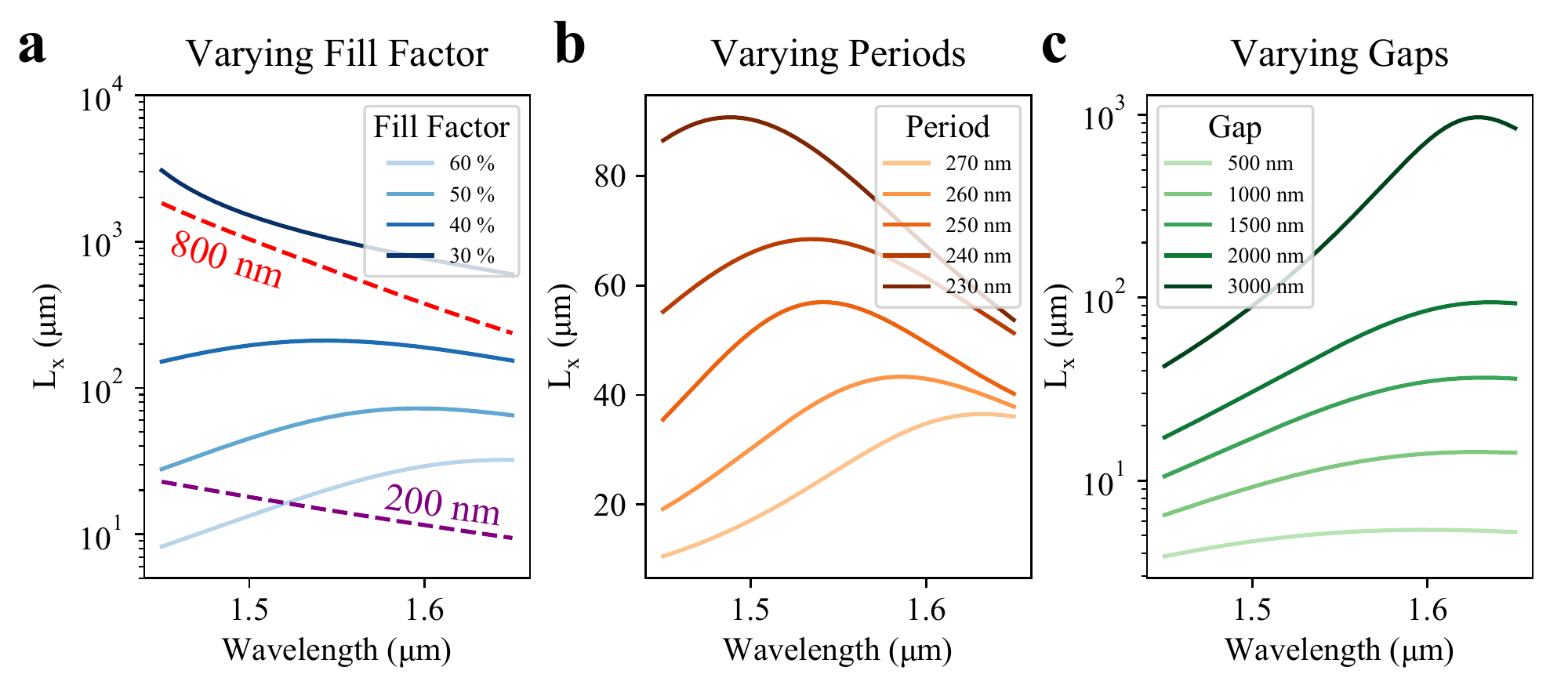}
    \caption{Dispersive plots representing the cross-over length for different varying parameters. (a) A fill factor sweep with period and gap fixed at $270$ nm and $1500$ nm, respectively. The reference lines indicate the cross-over lengths traditional directional coupler with gaps of  $200$ nm and $800$ nm.
    (b) A period sweep with fill factor and gap fixed at $60\%$ and $1500$ nm, respectively.
    (c) A coupling gap sweep with period and fill factor fixed at $270$ nm and $60\%$, respectively.}
    \label{fig: simulation}
\end{figure}  
We investigate the effect of different parameter variations of the large gap coupler. 
Fig. \ref{fig: simulation} shows the results of varying the fill factor, period and the gap between waveguides.
These parameter variations indicate the substantial tunability offered by two-directional \emph{e-skid}.
First, we selected the operating gap between the two waveguides to be  $1.44$ $\mu$m in order to stay consistent with the parallel \emph{e-skid} features ($\rho_{\parallel} = 60\%, \Lambda_{\parallel}=225$ nm, 6 layers deep results in a $1.44$ $\mu$m gap).
We designed these devices for manufacturing with the American Institute of Manufacturing (AIM) Photonics CMOS foundary Multi-Project Wafer (MPW) offering. 
For a photolithographic process like this one, we must take in to account the limitations of the processing, like feature size.
For example, many prior work designs with features smaller than $60$ nm would not resolve with CMOS processing compared to electron beam lithography.
We chose to design our devices with $\Lambda_{\bot} = 275$ nm to remain beneath the Bragg limit but maintain high manufacturing quality.
The parallel features were designed with $\Lambda_{\parallel} = 225$ nm.  
It should be noted that the parallel cladding structures are less challenging for a lithographic system because they are lines, not holes \cite{smith1998microlithography}.
We targeted  $\rho_{\bot} = 60 \%$ for the majority of our devices because we assumed that the features would be over etched, a common practice in SOI fabrication, so that the fill factor would decrease \cite{bogaerts2004basic}.

\section{Large Gap Directional Coupler Device Measurements \& Parameter Extraction}
\begin{figure}
    \centering
    \includegraphics[width = \textwidth]{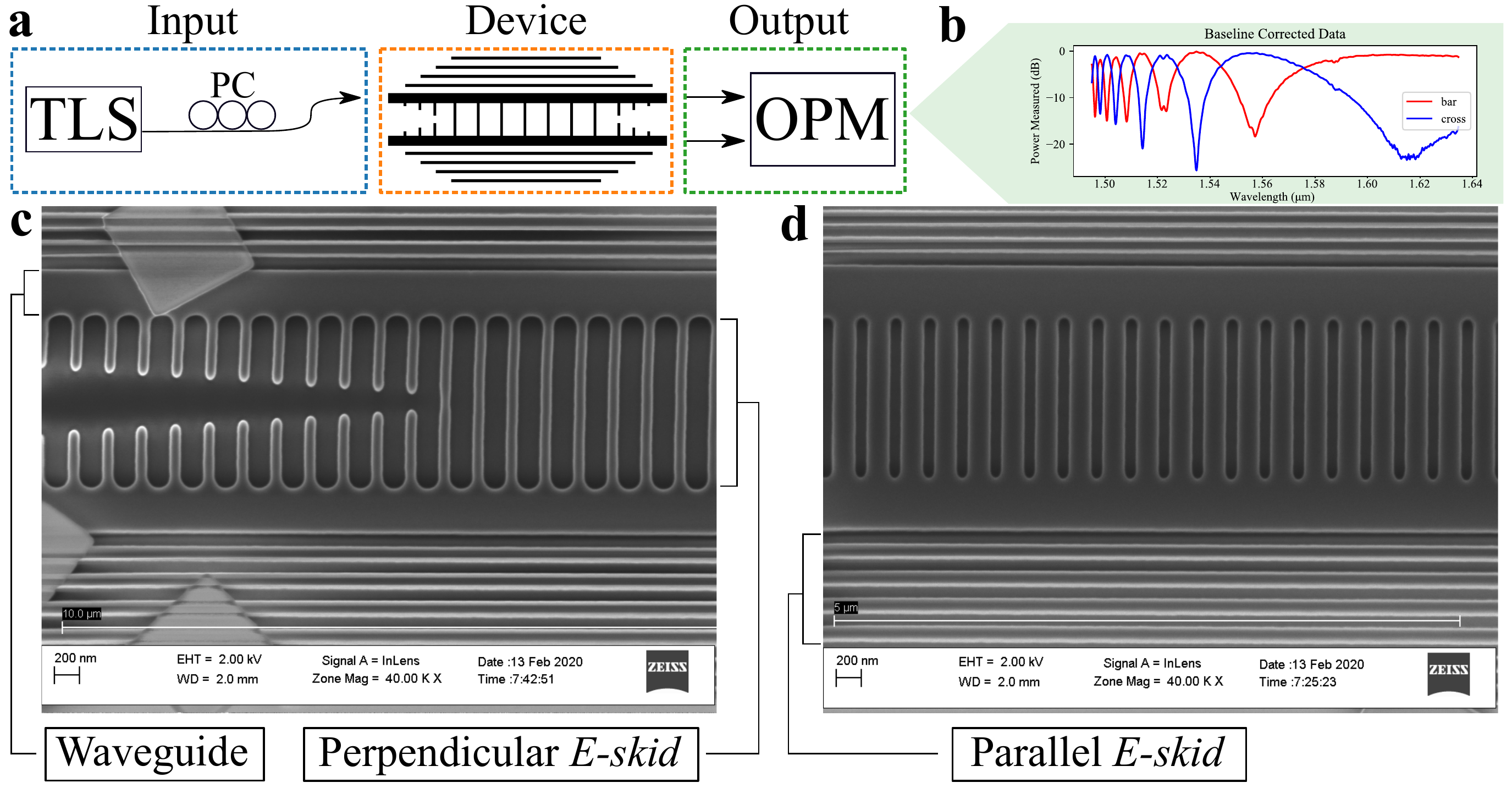}
    \caption{(a) Experimental setup, a tunable laser source (TLS) is connected to the device via a polarization controller (PC). The outputs of the device are then connected to an optical power meter (OPM).
    (b) An example spectrum from the measured data. 
    (c,d) An SEM image of the large gap directional coupler fabricated by AIM photonics, focused on the taper from the strip waveguides used to couple to optical fibers (c) and in the center of the coupler (d). The objects are debris from the oxide-release etch for the SEM.}
    \label{fig: SEM}
\end{figure}
\subsection{Experiment}
Our experimental setup is shown in Fig. \ref{fig: experiment} (a). We placed the chip on a mount in between two 3-axis stages with bare fiber on either side for coupling in and out. 
For the input, we connected the fiber to a tunable laser source (TLS), and a polarization controller (PC) to ensure TE polarization and we measured the output signal with an optical power meter (OPM).
The fibers were edge coupled to the chip, which routed the light through strip/wire waveguides to the devices. 
To transition from the strip waveguide mode to the \emph{e-skid}'s, we slowly introduced the parallel and perpendicular claddings as seen in Fig. \ref{fig: SEM} (c) and depicted in the schematic in Fig. \ref{fig: SEM} (a). 
We were careful to design simple tapers because when periodic, assymetric features are introduced there is a chance for radiative losses \cite{cheben2018subwavelength}.
In the future we will optimize these tapers to improve insertion loss, which were measured under $2$ dB per device.
\par 
We collected the transmission spectra (an example is shown in Fig. \ref{fig: SEM} (b)). The spectrum shows a characteristic ``chirp''-like behavior, which is due to the dispersive nature of the cross-over length. 
This is seen in Fig. \ref{fig: simulation} where, based on the perpendicular \emph{e-skid} parameters of the device, the cross-over length can exhibit a significant change with wavelength. 
This will result in a quickly oscillating output  (given by Eqs. (4) and (5)). With that said, using the experimental measurements in conjunction with the previously described models we were able to extract the parametric dependence of the coupler designs. 

\subsection{Parameter Extraction Method}
After extracting the transmission spectra from our fabricated devices, we used a dispersive model for extracting the cross-over length from the data.
Because inverse sine functions are multi-valued, we can not obtain $L_x$ directly from Eqs. \ref{eq: bar} or \ref{eq: cross}. 
We instead employed the behavioral model for characterizing directional couplers \cite{xing2017behavior}.
For the cross-over length, we are interested in the wavelength dependence, so we prepared the data by filtering the noise and normalizing the bar and cross measurements.
We used a polynomial expansion of the coupling coefficient coupled with a non-linear least squares (NLS) optimization algorithm to find the best fit for $L_x$ \cite{lumerical,scipy,numpy}.
Because many of our devices exhibits a strong ``chirp-like'' behaviour (Fig. \ref{fig: SEM} (b)), we used a third-order polynomial expression for $L_x$ to determine the best fit, such that
\begin{equation}\label{eq: poly}
L_x(\lambda) = L_{x,0} + L_{x,1}\lambda + L_{x,2}\lambda^2 + L_{x,3}\lambda^3,
\end{equation}
where the curve is characterized by fitting parameters $L_{x,0},L_{x,1},L_{x,2},L_{x,3}$ pertaining to the wavelength, $\lambda$.
The NLS optimization aimed to minimize the difference between the measured and theoretical spectra by adjusting the fit parameters in Eq. \ref{eq: poly}. 
\subsection{Experimental Results}
\begin{figure}
    \centering
    \includegraphics[width = \textwidth]{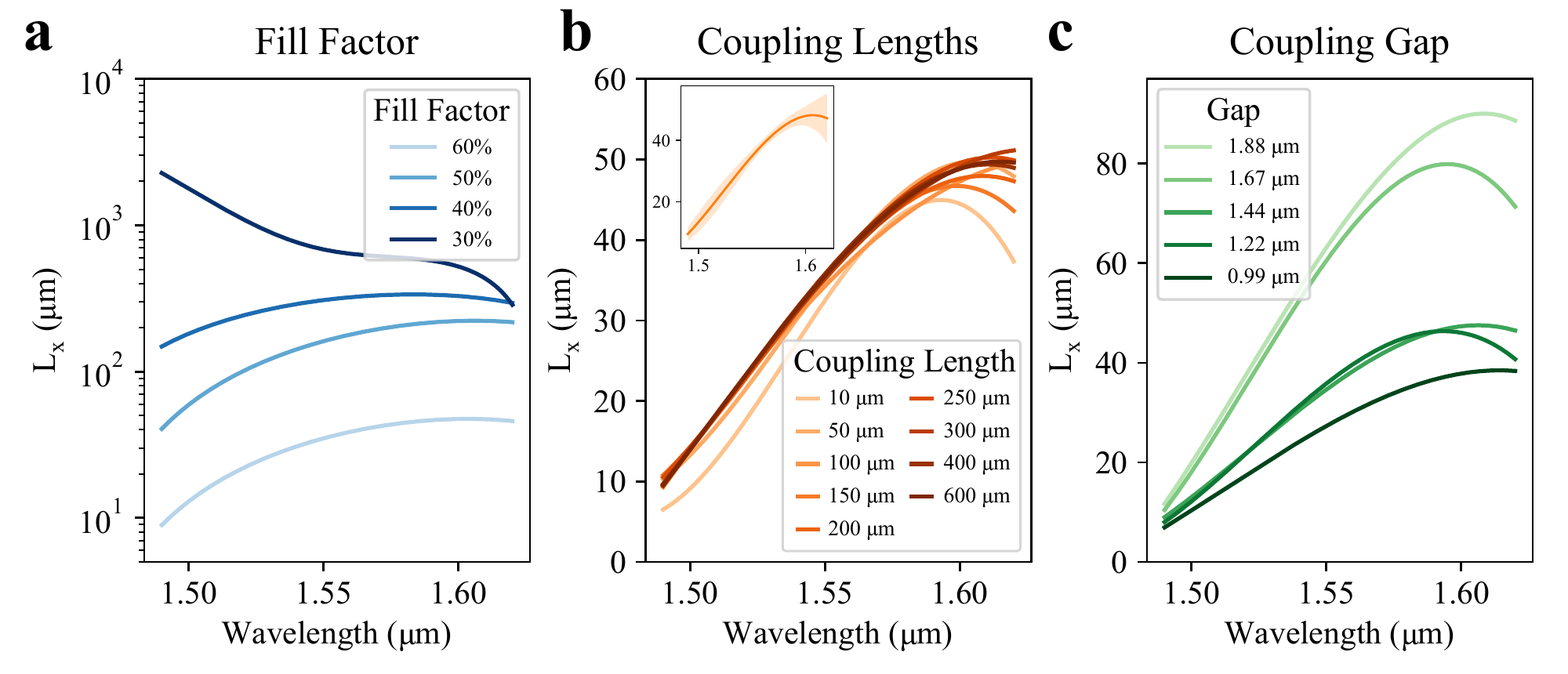}
    \caption{ (a) Experimental extraction of $L_x$ for the varying fill factors. 
    (b) Experimental extraction of $L_x$ for the varying coupling lengths, here we expect that the values will be similar.
    (c) Experimental extraction of $L_x$ for the varying coupling gaps, indicating an \emph{average} change of $L_x$, but less conclusive over the range.}
    \label{fig: experiment}
\end{figure}
In Fig. \ref{fig: experiment} (a), we show the devices dependence on fill factor variation.
Fill factors up to 60\% were successfully fabricated in the CMOS process and will be reported below, while fabrication specific optimization needs to go into higher fill factor devices.    
The fill factor variations resemble those from the simulations (Fig. \ref{fig: simulation} (a)), where higher fill factors increased $L_x$.
In Fig. \ref{fig: experiment} (b) we investigated coupler-length variation for a fixed fill factor (60\%) and gap ($1.44$ $\mu$m). 
We fit nine different directional couplers with the exact same parameters changing only the coupling length. According to Eqs. \ref{eq: bar} and \ref{eq: cross}, the length is independent of the coupling coefficient, $\kappa$ and therefore these couplers should exhibit identical $L_x$ measurements. 
There are manufacturing variations, measurement errors, and fitting errors that reveal themselves in Fig. \ref{fig: experiment} (b). The inset shows a 95 \% confidence interval for these nine couplers' $L_x$ extraction and showcase expected similar behavior for all of these couplers.
This truly highlights the utility of these devices, as they are able to couple fully in $\leq 50$ $\mu$m, even with a large gap of $1.44$ $\mu$m. 
Finally, Fig. \ref{fig: experiment} (c) shows the $L_x$ extraction for varying gaps. 
Even though there is a qualitative match between the simulations and experimental results, manufacturing variations account for the quantitative differences. 

\section{Future Work}\label{sec: discussion}

\begin{figure}
    \centering
    \includegraphics[width = \textwidth]{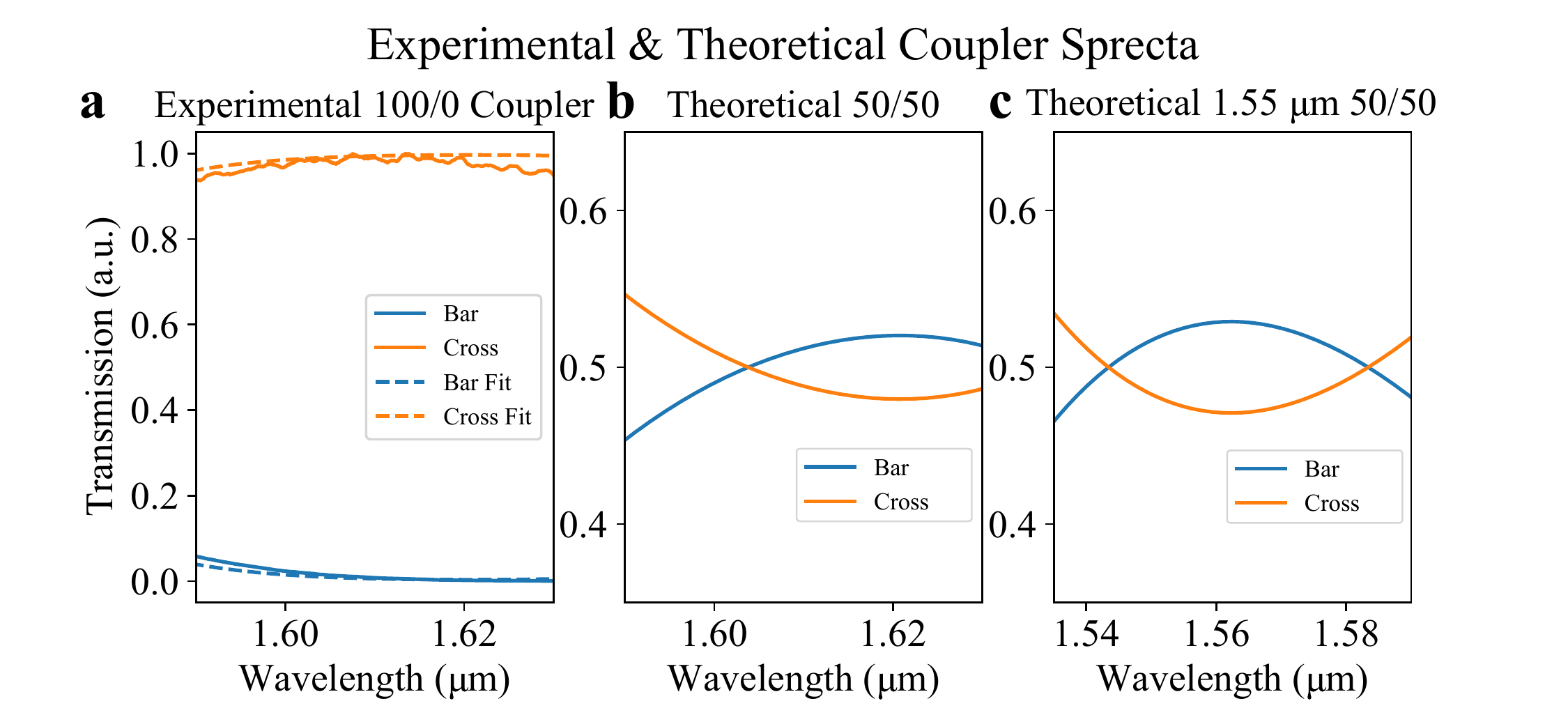}
    \caption{ 
    (a) Experimental and Fitting results for an $L_x$ Coupler ($100/0$) splitting ratio.
    (a) Based on the experimental results in (a), we propose a $50/50$ splitter operating from $1.58$ to $1.62$ $\mu$m.
    (b) Based on simulation results from Fig. \ref{fig: simulation} (a), we propose a $50/50$ splitter operating from $1.54$ to $1.58$ $\mu$m.}
    \label{fig: proposed}
\end{figure}

In this work we performed a comprehensive study of the parameter space of the large gap directional coupler using two-dimensional \emph{e-skid}. In the future, it will be desirable to realize devices with particular performance characteristics.  From our experimental data (Fig. \ref{fig: experiment}), we extracted that we are able to realize a large gap, bendless coupler that achieves 100\% coupling ($100/0$ splitting ratio) with a coupling length of $L=50$ $\mu$m as seen in Fig \ref{fig: proposed} (a), using $\rho_{\bot} = 60\%,\ \Lambda_{\bot} = 275 $ nm and coupling gap of $1.44$ $\mu$m.
We can take this to design a $50/50$ directional coupler.
Fig. \ref{fig: proposed} (b) shows the theoretical transmission spectrum of this device.
We set the coupling length $L = \operatorname{avg}(L_x)/2 = 23.5$ $\mu$m, and we see broadband behaviour of nearly $40$ nm. 
Additionally, we can slightly vary parameters to tune the device to a more desirable center wavelength.
For example, by reducing the period of the large gap directional coupler to  $\Lambda_{\bot} = 255$ nm, $\rho_{\bot} = 50\%$, and coupling gap of $1.44$ $\mu$m, the device's operating band can now be centered closer to 1.55 $\mu$m and achieves an even larger operating bandwidth of $>40$ nm (Fig. \ref{fig: proposed} (c)), which is sufficient for many applications.

\section{Conclusion}
We introduced the deterministic, targeted control of the evanescent wave in strip waveguides by employing \emph{e-skid} features in two directions. 
We designed and demonstrated large gap, bendless directional couplers through a CMOS photonic chip fabricated by AIM Photonics.  
All of the results were compared to simulations by extracting design parmaters using a NLS optimization technique coupled with a behavioral model of the directional coupler \cite{xing2017behavior}. 
With the parameter extraction, we show experimentally that \emph{e-skid} waveguides in general and the large gap, bendless directional coupler in particular are possible to realize in a CMOS platform. 
Moving forward, we can design large gap (>1$\mu$m) directional couplers that operate with bandwidths over $50$ nm.

\section*{Appendix A: Verifying Mode Conversion Efficiency of 1D \emph{e-skid} for CMOS Photonics}

\begin{figure}
    \centering
    \includegraphics[width = \textwidth]{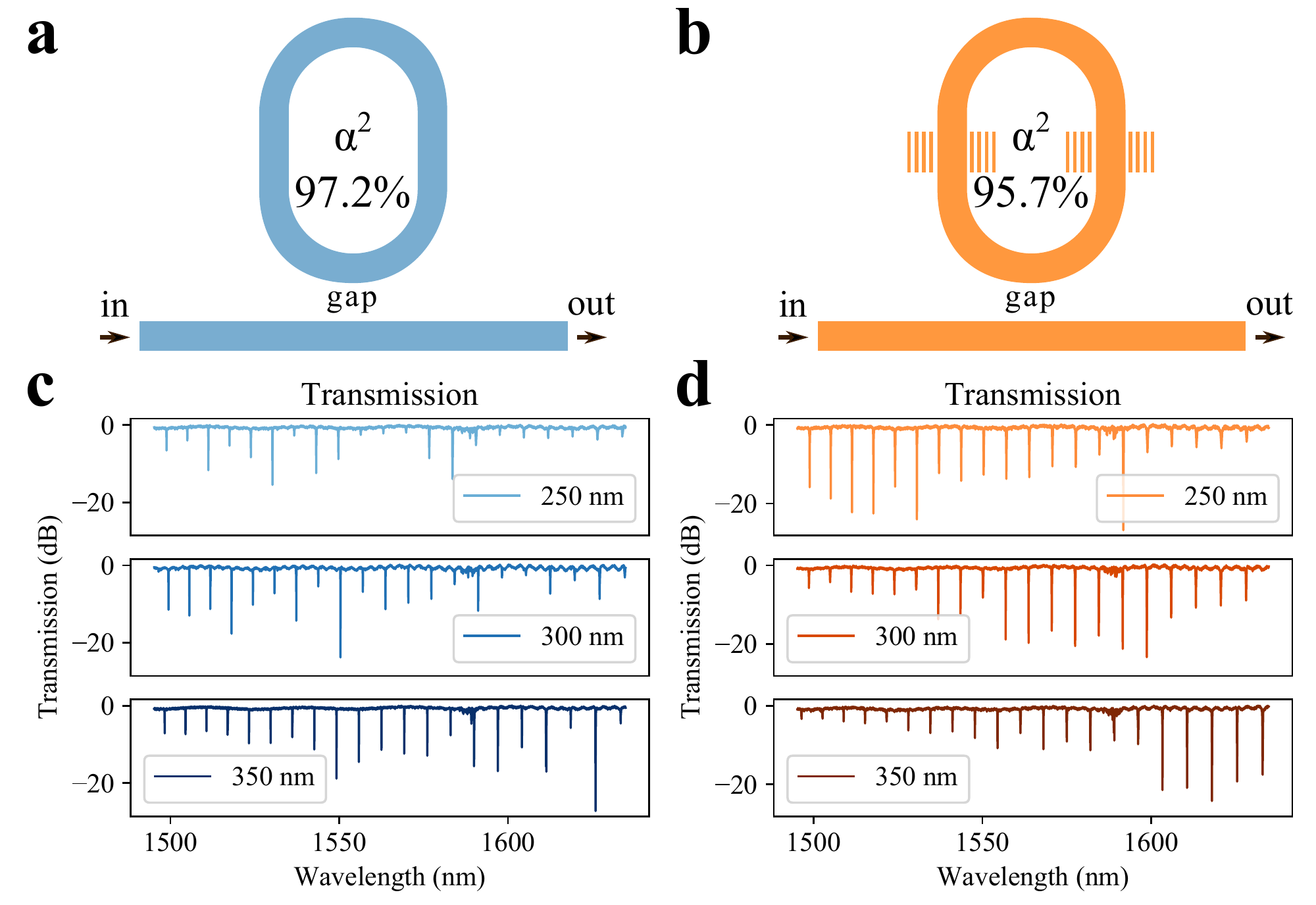}
    \caption{(a) A strip waveguide racetrack ring resonator. We measured an average loss into the ring of 97.2\% at the resonant peaks.
    (b) The transmission spectra for three equivalent strip waveguide racetrack ring resonators with varying coupling gaps of $250$, $300$ and $350$ nm. 
    (c) A strip waveguide racetrack ring resonator with two parallel cladding features for comparison with (a). The loss into the ring was measured at 95.7\% on average. 
    (d) Transmission spectra for the three rings of the same varying gaps as (b) with the addition of the features. }
    \label{fig: ring}
\end{figure}
We verify the conversion efficiency between strip waveguides and \emph{e-skid} waveguides by comparing the loss coefficients at the resonances of a racetrack resonator \cite{mckinnon2009extracting,han2016strip}.
We designed three strip waveguide racetrack resonators, Figure \ref{fig: ring} (a), with varying gaps, 
and then the exact same racetrack resonators in which we add two \emph{e-skid} features into the ring, Figure \ref{fig: ring} (c). 
The intent of this experiment is to quantify the additional loss created by these features, and thereby measure the mode conversion efficiency between strip and \emph{e-skid} waveguides \cite{jahani2018controlling}. 
A simple ring resonator (shown in Fig. \ref{fig: ring} (a)) can be paramaterized by two coefficients, $\tau$, the self-coupling coefficient that indicates how much light goes through the coupler, and $\alpha$, the loss coefficient which indicates how much light is lost into the ring.
We extract $\alpha$ and $\tau$ according to the method described by \cite{mckinnon2009extracting}, such that
\begin{align}
    \mathcal{F} &\equiv \frac{\Delta\lambda_{\text{FSR}}}{\Delta\lambda_{\text{FWHM}}},\\
    \mathcal{E} &\equiv \frac{T_{\text{MAX}}}{T_{\text{MIN}}},\\
    A &= \frac{\operatorname{cos}(\pi/\mathcal{F})}{1+\operatorname{sin}(\pi/\mathcal{F})},\\
    B &= 1-\left(1-\frac{\operatorname{cos}(\pi/\mathcal{F})}{1+\operatorname{cos}(\pi/\mathcal{F})}\right)\frac{1}{\mathcal{E}},\\
    (\alpha,\tau) &= \left[\frac{A}{B}\right]^{1/2}\pm\left[\frac{A}{B}-A\right]^{1/2}\label{eq: r1}.
\end{align}
The finesse, $\mathcal{F}$, is defined as the ratio between the free spectral range, $\Delta\lambda_{\text{FSR}}$, and the full width at half maximum, $\Delta\lambda_{\text{FWHM}}$, of each resonance. 
The exctinction ratio, $\mathcal{E}$, is defined as the ratio between the transmission maximum, $T_{\text{MAX}}$ , off resonance and the minimum, $T_{\text{MIN}}$ , at each resonance.
We can decouple $\alpha,\tau$ in Eq. \ref{eq: r1} using the method further discussed in \cite{mckinnon2009extracting}.
When we determine $\alpha$, we know that $\alpha^2$ indicates the percentage lost into the ring, which allows us to compare these two different resonators. 
This resulted in average values of $\alpha^2 = 97.2\%$ and $\alpha^2 = 95.7\%$ for the two ring types (Fig. \ref{fig: ring} (a,c)), respectively. This leads to a mode conversion efficiency of $99.6$\%.
\section*{Funding}
National Science Foundation (Award\# 1810282),  Air Force Research Laboratory (FA8650-15-2-5220 \& FA8750-16-2-0140)

\section*{Acknowledgements}
This material is based upon work supported by the National Science Foundation under Grant No. 1810282 and AFRL awards (FA8650-15-2-5220 \& FA8750-16-2-0140). Any opinions, findings, and conclusions or recommendations expressed in this material are those of the author(s) and do not necessarily reflect the views of the National Science Foundation. The U.S. Government is authorized to reproduce and distribute reprints for Governmental purposes notwithstanding any copyright notation thereon. The views and conclusions contained herein are those of the authors and should not be interpreted as necessarily representing the official policies or endorsements, either expressed or implied, of Air Force Research Laboratory or the U.S. Government. M.v.N. and S.F.P would like to acknowledge Navin B. Lingaraju for sparking a collaboration between RIT and Purdue.

\section*{Disclosures} 
The authors declare that there are no conflicts of interest related to this article.

\bibliography{bib}

\end{document}